\documentclass[aps, floatfix]{revtex4}
\usepackage{graphics}
\usepackage{epsfig}
\usepackage{subfigure}
\usepackage{amsmath, amsfonts, amssymb, graphicx, times}

\begin{document}

\title{Exp-function
method for solving the Burgers-Fisher equation with variable
coefficients}

\author{Bo-Kui Chen$^1$}\email{chenssx@mail.ustc.edu.cn}
\author{Yang Li$^1$}
\author{Han-Lin Chen$^2$}
\author{Bing-Hong Wang$^1$, $^3$}

\affiliation{$^{1}$Department of Modern Physics, University of
Science and Technology of China, Hefei 230026, China\\$^{2}$School
of Science, Southwest University of Science and Technology, Mianyang
621010, China\\$^{3}$The Research Center for Complex System Science,
University of Shanghai for Science and Technology and Shanghai
Academy of System Science, Shanghai 200093 China }

\begin{abstract}
In this paper, the exp-function method with the aid of symbolic
computational system is used to obtain generalized travelling wave
solutions of a Burgers-Fisher equation with variable coefficients.
It is shown that the exp-function method, with the help of symbolic
computation, provides a straightforward and powerful mathematical
tool to solve the nonlinear evolution equation
with variable coefficients in mathematical physics.
\end{abstract}

\maketitle



\section{Introduction}
The investigation of exact solutions of nonlinear evolution
equations(NLEEs) plays an important role in the study of nonlinear
physics phenomena. The importance of obtaining the exact solutions
of these nonlinear equations, if available, will facilitate the
verification of numerical solvers and aids in the stability analysis
of solutions. In the past several decades, many effective methods
for obtaining exact solutions of NLEEs have been presented, such as
the tanh-function method \cite{1, 2, 3}, extended tanh method
\cite{4, 5}, F-expansion method \cite{6, 7}, sine-cosine method
\cite{8, 9, 10}, Jacobian elliptic function method
\cite{11, 12, 13, 14}, homotopy perturbation method \cite{15, 16, 17, 18},
variational iteration method \cite{19, 20} and Adomian method
\cite{21, 22, 23} and so on.\par Recently, He and Wu \cite{24}
proposed a straightforward and concise method, called Exp-function
method, to obtain generalized solitary solutions and periodic
solutions. Applications of the Exp-function method can be found in
\cite{25, 26, 27} for solving nonlinear evolution equations arising in
mathematical physics. The solution procedure of this method, with
the aid of Maple, is of utter simplicity and this method can easily
extended to other kinds of nonlinear evolution equations.\par

The present Letter is motivated by the desire to extend the
exp-function method to the general types of Burgers-Fisher equation
with variable coefficients, which reads:
\begin{equation}\label{BF}
 u_{t}-u_{xx}+\alpha(t) uu_{x}=\beta(t){u}(1-u),
\end{equation}
where $\alpha(t), \beta(t)$ are arbitrary functions of $t$.\par

The Burgers-Fisher equation has a wide rang of applications in
plasma physics, fluid physics, capillary-gravity waves, nonlinear
optics and chemical physics. When
$\alpha(t)=0$, $\beta(t)$ is a arbitrary constant, Eq.(\ref{BF})
turns to Fisher equation,
\begin{equation}\label{F}
 u_{t}-u_{xx}=\beta{u}(1-u),
\end{equation}
Kolmogorov, Petrovskii, and Piskunov studied this equation in
\cite{28}They showed if initial datum satisfies some conditions then
the solution of Eq.(\ref{F}) approaches a travelling wave of speed
$C_{0}=2$. Exact solution of Eq.(\ref{F}) was found by Ablowitz and
Zeppetella in \cite{29} at $C_{0}=\pm\frac{5}{\sqrt{6}}$. When
$\beta(t)=0$, $\alpha(t)$ is a arbitrary constant, Eq.(\ref{BF}) turns
to Burgers equation,
\begin{equation}\label{B}
 u_{t}-u_{xx}+\alpha uu_{x}=0,
\end{equation}
which is used to describe the spread of sound wave in the medium with
viscidity and heat exchange if we do not consider the medium's
frequently dispersive character and the slack comfort process, at
the same time, the Burgers equations with variable coefficient can
be used to describe the cylinder and spherical wave in these
questions such as overfall, traffic flow model and so on.\par

Therefore, there are important theoretic and factual value for us to
look for the exact solutions of Eq.(\ref{BF}).

\section{The Exp-function method}
We now present briefly the main steps of the Exp-function method
that will be applied.  A traveling wave transformation
$u=u(\xi)$, $\xi=kx+wt$ converts a partial differential equation
\begin{equation}\label{ef1}
\Psi(u, u_{t}, u_{x}, u_{xx}, u_{tt}, \cdots)=0,
\end{equation}
into an ordinary differential equation

\begin{equation}\label{ef2}
\Phi(u, wu', ku', k^2u'', w^2u'', kwu'', \cdots)=0,
\end{equation}\par

The Exp-function method is based on the assumption that traveling
wave solutions can be expressed in the following from \cite{24}
\begin{equation}\label{ef3}
u(\xi)=\frac{\sum\nolimits_{n=-c}^{d}a_{n}\exp(n\xi)}{\sum\nolimits_{m=-p}^{q}b_{m}\exp(m\xi)},
\end{equation}
where $c, d, p$ and $q$ are positive integers which are unknown to be
further determined, $a_{n}$ and $b_{m}$ are unknown constants. To
determine the values of $c$ and $p$, and the value of $d$ and $q$,
we balance the linear term of highest order in Eq. (\ref{ef2}) with
the lowest order nonlinear term, respectively.

\section{Application to the Burgers-Fisher equation}
In order to obtain the solution of Eq. (\ref{BF}), we consider the
transformation
\begin{equation}\label{bh}
u=u(\xi), \quad \xi=kx+\int \tau(t)dt,
\end{equation}
where $k$ is a constant, $\tau(t)$is an integrable function of $t$
to be determined later, then Eq. (\ref{BF}) becomes an ordinary
differential equation, which reads
\begin{equation}\label{OBF}
\tau(t)u'+k\alpha(t){uu'}-k^2u''-\beta(t)u(1-u)=0,
\end{equation}
where prime denotes the differential with respect to $\xi$.\par

 According to the Exp-function
method, we assume that the solution of Eq. (\ref{OBF}) can be
expressed in the form

\begin{equation}\label{SOBF}
u(\xi)=\frac{\sum\nolimits_{n=-c}^{d}a_{n}\exp(n\xi)}{\sum\nolimits_{m=-p}^{q}b_{m}\exp(m\xi)}=\frac{a_{-c}\exp(-c\xi)+\cdots+a_{d}\exp(d\xi)}
{b_{-p}\exp(-p\xi)+\cdots+b_{q}\exp(q\xi)},
\end{equation}
where $c, d, p$ and $q$ are positive integers which are unknown to be
further determined, $a_{n}$ and $b_{m}$ are unknown constants.\par

In order to determine values of $c$ and $p$, we balance the linear
term of highest order in Eq.     (\ref{OBF})with the highest order
nonlinear term, and the linear term of lowest order in Eq.
(\ref{OBF}) with the lowest order nonlinear term, respectively. By
simple calculation, we have
\begin{equation}\label{hl}
u''(\xi)=\frac{h_{1}\exp[(d+3q)\xi]+\cdots}{h_{2}\exp(4q\xi)+\cdots},
\end{equation}
and
\begin{equation}\label{nhl}
u(\xi)u'(\xi)=\frac{h_{3}\exp(2d+q)\xi+\cdots}{h_{4}\exp(3q\xi)+\cdots}=\frac{h_{3}\exp(2d+2q)\xi+\cdots}{h_{4}\exp(4q\xi)+\cdots},
\end{equation}
where $h_{i}$ are determined coefficients only for
simplicity.
Balancing highest order of Exp-function in Eq.
(\ref{hl})and      (\ref{nhl}), we have
\begin{equation}
d+3q=2d+2q, \quad d=q.
\end{equation}\par
Similarly to determine values of $c$ and $p$, we balance the linear
term of lowest order in Eq.     (\ref{OBF})
\begin{equation}\label{ol}
u''(\xi)=\frac{\cdots+s_{1}\exp[-(c+3p)\xi]}{\cdots+s_{2}\exp(-4p\xi)},
\end{equation}
and
\begin{equation}\label{nol}
u(\xi)u'(\xi)=\frac{s_{3}\exp[-(2c+p)]\xi+\cdots}{\cdots+s_{4}\exp(-3p\xi)}=\frac{\cdots+s_{3}\exp[-(2c+2p)]\xi}{\cdots+s_{4}\exp(-4p\xi)},
\end{equation}
where $s_{i}$ are determined coefficients only for simplicity.
Balancing highest order of Exp-function in Eq. (\ref{ol})and
(\ref{nol}), we have
\begin{equation}
c+3p=2c+2p, \quad c=p.
\end{equation}

we can freely choose the values of $c$ and $d$, but the final
solution does not strongly depends upon the choice of values of $c$
and $d$[26].  For simplicity, we set $b_{1}=1$, $p=c=1$ and $d=q=1$
Eq. (\ref{SOBF})becomes
\begin{equation}\label{SOBF1}
u(\xi)=\frac{a_{1}\exp(\xi)+a_{0}+a_{-1}\exp(-\xi)}{\exp(\xi)+b_{0}+b_{-1}\exp(-\xi)},
\end{equation}
Substituting Eq.     (\ref{SOBF1})into Eq.     (\ref{OBF}), and with
the help of Maple, we have
\begin{equation}
\frac{1}{A}[C_{3}\exp(3\xi)+C_{2}\exp(2\xi)+C_{1}\exp(\xi)+C_{0}+C_{-1}\exp(-\xi)+C_{-2}\exp(-2\xi)+C_{-3}\exp(-3\xi)]=0,
\end{equation}
where
\begin{eqnarray*}
A&=&(\exp(\xi)+b_{0}+b_{-1}\exp(-\xi))^{3};\\
C_{3}&=&-a_{{1}}\beta(t)+{a^{2}_{{1}}}\beta(t);\\
C_{2}&=&-2\, a_{{1}}b_{{0}}\beta(t)-ka_{{1}}a_{{0}}\alpha(t)+{a^{2}_{{1}}};
b_{{0}}\beta(t)+a_{{1}}b_{{0}}\tau(t)+k{a^{2}_{{1}}}b_{{0}}\alpha(t)-a_{{0}}\tau(t)-a_{{0}}\beta(t)+{k}^{2}a_{{1}}b_{{0}}-{k}^{2}a_{{0}}+2\, a_{{1}}a_{{0}}\beta(t);\\
C_{1}&=&{a^{2}_{{0}}}\beta(t)-2\, ka_{{1}}a_{{-1}}\alpha(t)-a_{{0}}b_{{0}}\tau(t)+2\, a_{{0}}a_{{1}}b_{{0}}\beta(t)-{k}^{2}a_{{1}}{b^{2}_{{0}}}-k{a^{2}_{{0}}}\alpha(t)+2k{a^{2}_{{1}}}b_{{-1}}\alpha(t)-a_{{-1}}\beta(t)-2a_{{-1}}\tau(t);\\
&&+a_{{1}}{b^{2}_{{0}}}\tau(t)-2\, a_{{0}}b_{{0}}\beta(t)+2\, a_{{1}}a_{{-1}}\beta(t)+{k}^{2}a_{{0}}b_{{0}}
-a_{{1}}{b^{2}_{{0}}}\beta(t)-4\, {k}^{2}a_{{-1}}+2a_{{1}}b_{{-1}}\tau(t)+4{k}^{2}a_{{1}}b_{{-1}}\\
&&-2a_{{1}}b_{{-1}}\beta(t)+{a^{2}_{{1}}}b_{{-1}}\beta(t)+ka_{{0}}a_{{1}}b_{{0}}\alpha(t);\\
C_{0}&=&6{k}^{2}a_{{0}}b_{{-1}}+2a_{{0}}a_{{-1}}\beta(t)-2a_{{1}}b_{{0}}b_{{-1}}\beta(t)-2a_{{-1}}b_{{0}}\beta(t)-3a_{{-1}}b_{{0}}\tau(t)-3{k}^{2}a_{{-1}}b_{{0}}+3a_{{1}}b_{{0}}b_{{-1}}\tau(t)-3{k}^{2}a_{{1}}b_{{0}}b_{{-1}};\\
&&+2a_{{1}}a_{{2}};
b_{{0}}\beta(t)+2a_{{1}}a_{{0}}b_{{-1}}\beta(t)-3ka_{{0}}a_{{-1}}\alpha(t)+3ka_{{1}}a_{{0}}b_{{-1}}\alpha(t)-2a_{{0}}b_{{-1}}\beta(t)-a_{{0}}{b^{2}_{{0}}}\beta(t)+{a^{2}_{{0}}}b_{{0}}\beta(t);\\
C_{-1}&=&-2k{a^{2}_{{-1}}}\alpha(t)-a_{{1}}{b^{2}_{{-1}}}\beta(t)+{a_{{2}};
}^{2}\beta(t)+{a^{2}_{{0}}}b_{{-1}}\beta(t)+2ka_{{-1}}a_{{1}}b_{{-1}}\alpha(t)+2a_{{1}}{b^{2}_{{-1}}}\tau(t)-a_{{-1}}{b^{2}_{{0}}}\tau(t)
+2a_{{0}}a_{{-1}}b_{{0}}\beta(t)\\
&&-a_{{-1}}{b^{2}_{{0}}}\beta(t)+{k}^{2}a_{{0}}b_{{-1}}b_{{0}}+2a_{{-1}}a_{{1}}b_{{-1}}\beta(t)-2a_{{0}}b_{{-1}}b_{{0}}\beta(t)
-4{k}^{2}a_{{1}}{b^{2}_{{-1}}}+a_{{0}}b_{{-1}}b_{{0}}\tau(t)-2a_{{-1}}b_{{-1}}\beta(t)\\
&&-ka_{{0}}a_{{-1}}b_{{0}}\alpha(t)-2a_{{-1}}b_{{-1}}\tau(t)
-{k}^{2}a_{{-1}}{b^{2}_{{0}}}+4{k}^{2}a_{{-1}}b_{{-1}}+k{a^{2}_{{0}}}b_{{-1}}\alpha(t);\\
C_{-2}&=&-2a_{{-1}}b_{{0}}b_{{-1}}\beta(t)-{k}^{2}a_{{0}}{b^{2}_{{-1}}}-a_{{0}}{b^{2}_{{-1}}}\beta(t)-k{a^{2}_{{-1}}}b_{{0}}\alpha(t);
+{a^{2}_{{-1}}}b_{{0}}\beta(t)+{k}^{2}a_{{-1}}b_{{0}}b_{{-1}}+a_{{0}}{b^{2}_{{-1}}}\tau(t)\\
&&+ka_{{2}}a_{{0}}b_{{-1}}\alpha(t)
-a_{{-1}}b_{{0}}b_{{-1}}\tau(t)+2a_{{-1}}a_{{0}}b_{{2}}\beta(t);\\
C_{-3}&=&{a^{2}_{{-1}}}b_{{-1}}\beta(t)-a_{{-1}}{b^{2}_{{-1}}}\beta(t).\\
\end{eqnarray*}\par
Equating to zero the coefficients of all powers of $\exp(n\xi)$ yields a set of algebraic equations for
$a_{1}, a_{0}, a_{-1}, b_{-1} $, $b_{0}, b_{1}, k, \tau(t), \alpha(t), \beta(t)$
. Solving the system of algebraic equations with the aid of Maple, we
obtain:\par \textbf{case1.}
\begin{equation}\label{case1}
\tau(t)=-k^2-\beta(t), \quad \alpha(t)=\frac{\beta(t)}{k}, \quad
a_{1}=0, \quad a_{0}=a_{0}, \quad a_{-1}=a_{0}b_{0}, \quad
b_{0}=b_{0}, \quad b_{-1}=0.
\end{equation}\par

\textbf{case2.}
\begin{equation}\label{case2}
\tau(t)=2k^2+\frac{1}{2}\beta(t), \quad \alpha(t)=-4k, \quad
a_{1}=1, \quad a_{0}=0, \quad a_{-1}=0, \quad b_{0}=0, \quad
b_{-1}=b_{1}.
\end{equation}\par

\textbf{case3.}
\begin{equation}\label{case3}
\tau(t)=-k^2-\beta(t), \quad \alpha(t)=2k, \quad a_{1}=0, \quad
a_{0}=\frac{b_{0}+\sqrt{b^{2}_{0}-4b_{-1}}}{2}, \quad
a_{-1}=b_{-1}, \quad b_{0}=b_{0}, \quad b_{-1}=b_{-1}.
\end{equation}\par

\textbf{case4.}
\begin{equation}\label{case4}
\tau(t)=k^2+\beta(t), \quad \alpha(t)=-2k, \quad a_{1}=1, \quad
a_{0}=\frac{b_{0}-\sqrt{b_{0}^{2}-4b_{-1}}}{2}, \quad a_{-1}=0, \quad
b_{0}=b_{0}, \quad b_{-1}=b_{-1}.
\end{equation}\par

\textbf{case5.}
\begin{equation}\label{case5}
\tau(t)=-k^2, \quad \alpha(t)=\frac{\beta(t)}{k}, \quad a_{1}=1, \quad
a_{0}=a_{0}, \quad a_{-1}=-b^{2}_{0}+a_{0}b_{0}, \quad
b_{0}=b_{0}, \quad b_{-1}=0.
\end{equation}\par

\textbf{case6.}
\begin{equation}\label{case61}
\tau(t)=\frac{k^2(12\sqrt{2b_{-1}}a_{0}^{4}+7\sqrt{2}b_{-1}^{\frac{5}{2}}+40\sqrt{2}b_{-1}^{\frac{3}{2}}a_{0}^{2}+45a_{0}^{3}b_{-1}+2a_{0}^{5}+37a_{0})}
{10\sqrt{2}a_{0}^{2}b_{-1}^{\frac{3}{2}}+\sqrt{2}b_{-1}^{\frac{5}{2}}+6\sqrt{2b_{-1}}a_{0}^{4}+2a_{0}^{5}+7a_{0}b_{-1}^{2}+15a_{0}^{3}b_{-1}}, \quad
\alpha(t)=-2k, \quad  b_{-1}=b_{-1},
\end{equation}
\begin{equation}\label{case62}
\beta(t)=\frac{6k^2(2a_{0}b_{-1}^{\frac{3}{2}}\sqrt{2}+\sqrt{2b_{-1}}a_{0}^{3}+3a_{0})}
{4\sqrt{2b_{-1}}a_{0}^{3}+3\sqrt{2}a_{0}b_{-1}^{\frac{3}{2}}+7a_{0}^{2}b_{-1}+2a_{0}^{4}+b_{-1}^{2}}, \quad
a_{1}=1, \quad a_{0}=a_{0}, \quad a_{-1}=0, \quad
b_{0}=-\sqrt{2b_{-1}}.
\end{equation}\\[0.4cm]

Substituting Eqs. (\ref{case1})-(\ref{case62}) into Eq.
(\ref{SOBF1}) yields

\begin{equation}\label{solution1}
u_{1}(x, t)=\frac{a_{0}+a_{0}b_{0}\exp\left[-kx+\int{k^{2}+\beta(t)}dt\right]}
{\exp\left[kx-\int{k^{2}+\beta(t)}dt\right]+b_{0}},
\end{equation}

\begin{equation}\label{solution2}
u_{2}(x, t)=\frac{\exp\left[kx+\int{2k^2+\frac{\beta(t)}{2}}dt\right]}
{\exp\left[kx+\int{2k^2+\frac{\beta(t)}{2}}dt\right]+b_{-1}\exp\left[-kx-\int{2k^2+\frac{\beta(t)}{2}}dt\right]},
\end{equation}

\begin{equation}\label{solution3}
u_{3}(x, t)=\frac{\frac{b_{0}+\sqrt{b_{0}^{2}-4b_{-1}}}{2}+b_{-1}\exp\left[-kx+\int{k^{2}+\beta(t)}dt\right]}
{\exp\left[kx-\int{k^{2}+\beta(t)}dt\right]+b_{0}+b_{-1}\exp\left[-kx+\int{k^{2}+\beta(t)}dt\right]},
\end{equation}

\begin{equation}\label{solution4}
u_{4}(x, t)=\frac{\exp\left[kx+\int{k^2+\beta(t)}dt\right]+\frac{b_{0}-\sqrt{b_{0}^{2}-4b_{-1}}}{2}}
{\exp\left[kx+\int{k^2+\beta(t)}dt\right]+b_{0}+b_{-1}\exp\left[-kx-\int{k^2+\beta(t)}dt\right]},
\end{equation}

\begin{equation}\label{solution5}
u_{5}(x, t)=\frac{\exp(kx-k^{2}t)+a_{0}+(-b_{0}^{2}+a_{0}b_{0})\exp(-kx+k^{2}t)}
{\exp(kx-k^{2}t)+b_{0}},
\end{equation}

\begin{equation}\label{solution6}
u_{6}(x, t)=\frac{\exp(kx+\int{\tau(t)}dt)+a_{0}}{\exp(kx+\int{\tau(t)}dt)-\sqrt{2b_{-1}}+b_{-1}\exp(-kx-\int{\tau(t)}dt)},
\end{equation}
where
$\tau(t)=\frac{k^2(12\sqrt{2b_{-1}}a_{0}^{4}+7\sqrt{2}b_{-1}^{\frac{5}{2}}+40\sqrt{2}b_{-1}^{\frac{3}{2}}a_{0}^{2}+45a_{0}^{3}b_{-1}+2a_{0}^{5}+37a_{0})}
{10\sqrt{2}a_{0}^{2}b_{-1}^{\frac{3}{2}}+\sqrt{2}b_{-1}^{\frac{5}{2}}+6\sqrt{2b_{-1}}a_{0}^{4}+2a_{0}^{5}+7a_{0}b_{-1}^{2}+15a_{0}^{3}b_{-1}}$.
As we will show later, these solution have included all the solutions obtained in \cite{30} using the first integral method.

\section{Some discussions about the solutions}
\begin{enumerate}
  \item[(1)] If we take $b_{0}=0$ in Eq. (\ref{solution1}), we have
  \begin{equation}\label{DS1}
    u_{11}(x, t)=a_{0}\exp\left[-kx+\int{k^{2}+\beta(t)}dt\right].
  \end{equation}

  \item[(2)]If we take $\alpha(t)=a$ is a constant, $b_{-1}=1$ and
  $\beta(t)=\frac{2ac-a^2}{4}$ in Eq. (\ref{solution2}), where $c$ is
  a arbitrary constant. Then we have,
  \begin{equation}\label{DS2}
    u_{21}(x, t)=\frac{1}{2}-\frac{1}{2}\tanh\left[\frac{a}{4}(x-ct)\right].
  \end{equation}
  This is the solution (36) obtained in \cite{30}. Other solutions in \cite{30} also included in our
  solution (25) - (30).
  \item[(3)] If we take $b_{0}=4$, $b_{-1}=1$ and $k=1$ in Eq.(\ref{solution3}), we
  have,
\begin{equation}\label{DS3}
u_{31}(x, t)=\frac{-1+(2+2\sqrt{3})\csc\left[x-\int{1+\beta(t)}dt\right]+\coth\left[x-\int{1+\beta(t)}dt\right]}
{4\csc\left[x-\int{1+\beta(t)}dt\right]+2\coth\left[x-\int{1+\beta(t)}dt\right]}.
\end{equation}

  \item[(4)] If we take $b_{0}=0$, $b_{-1}=-5$ and $k=1$ in Eq. (\ref{solution4}). We get,

\begin{equation}\label{DS4}
u_{41}(x, t)=\frac{\cosh\left[x+\int{1+\beta(t)}dt\right]+\sinh\left[x+\int{1+\beta(t)}dt\right]-\sqrt{5}}
{-4\cosh\left[x+\int{1+\beta(t)}dt\right]+6\sinh\left[x+\int{1+\beta(t)}dt\right]}.
\end{equation}

\item[(5)] If we take $b_{0}=2$ and $a_{0}=\frac{3}{2}$, we have
\begin{equation}\label{DS5}
u_{51}=\frac{2\tanh(kx-k^2t)+\frac{3}{2}\sec(kx-k^2t)}{1+\tanh(kx-k^2t)+2\sec(kx-k^2t)},
\end{equation}

\item[(6)] If we take $a_{0}=0$ in Eq.(\ref{solution6}), we have
\begin{equation}\label{DS6}
u_{61}(x, t)=\frac{\exp(kx+7k^2t)}{\exp(kx+7k^2t)-\sqrt{2b_{-1}}+b_{-1}\exp(-kx-7k^2t)},
\end{equation}

\end{enumerate}
These solutions for local area structure are shown in Figs. 1-3.

\begin{figure}
\scalebox{0.6}[0.4]{\includegraphics{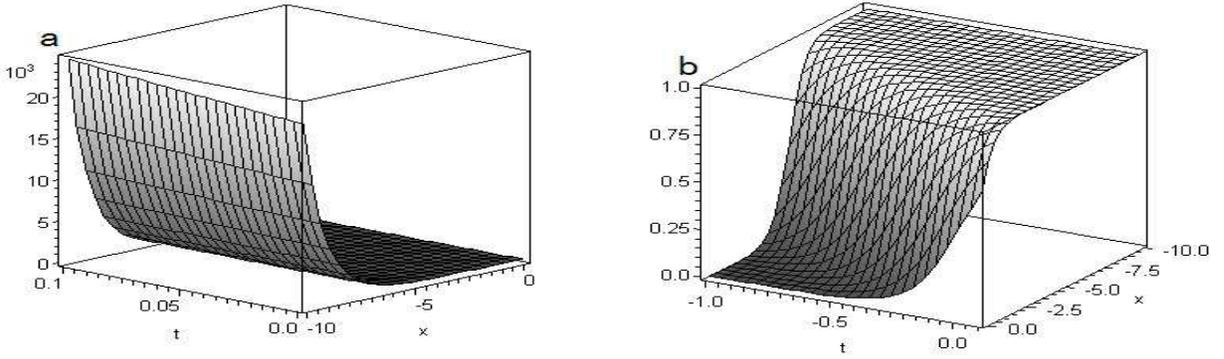}}
\caption{\label{fig:12}(a) Solution of Eq.(\ref{DS1}) with
$a_{0}=1$, $k=1$ and $\beta(t)=t$. \qquad (b) Solution of
Eq.(\ref{DS2}) with $a=4$ and $c=5$.}
\end{figure}
\begin{figure}
\scalebox{0.8}[0.6]{\includegraphics{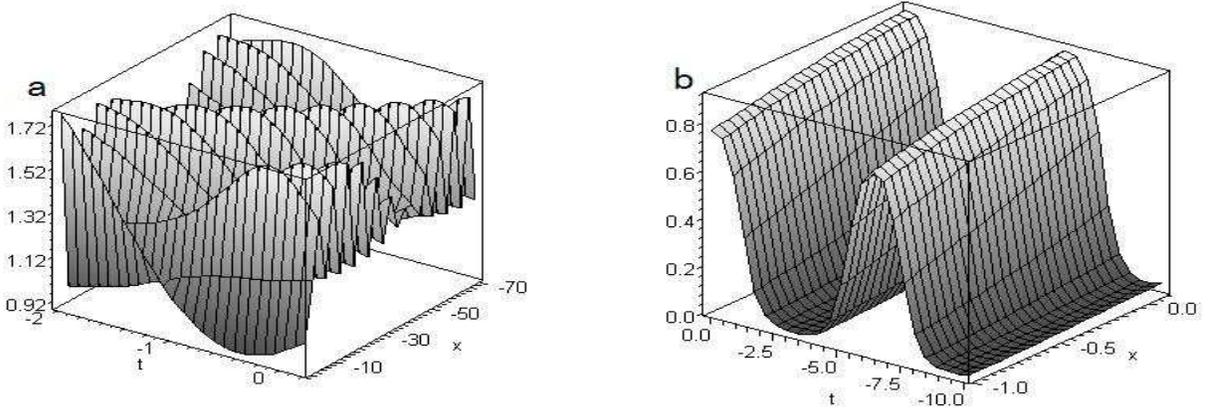}}
\caption{\label{fig:34}(a) Solution of Eq.(\ref{DS3}) with
 $\beta(t)=\cos{t}$. \qquad(b) Solution of Eq.(\ref{DS4})
with $\beta(t)=-1+3\sin{t}$.}
\end{figure}
\begin{figure}
\scalebox{0.6}[0.4]{\includegraphics{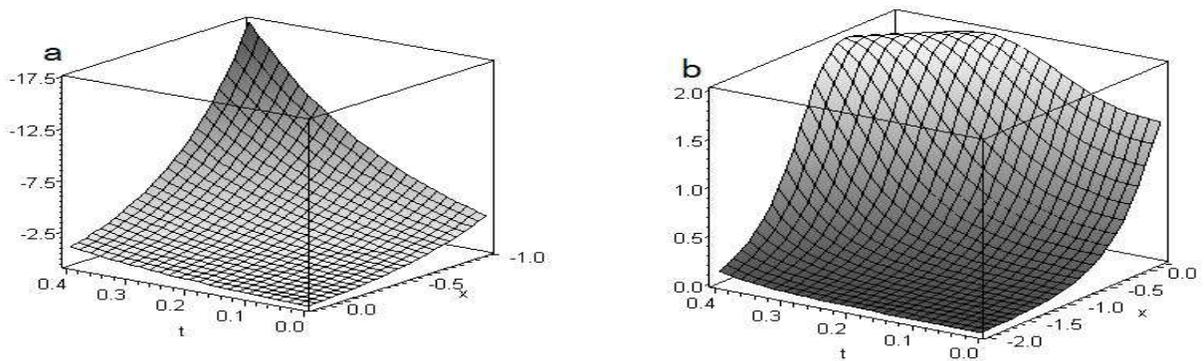}}
\caption{\label{fig:56}(a) Solution of Eq.(\ref{DS5}) with
 $k=2$.\qquad (b) Solution of Eq.(\ref{DS6})
with $k=1$ and $b_{-1}=2$.}
\end{figure}

\section{conclusion}
The Burgers-Fisher equation with variable coefficients is
investigated by Exp-function method. The generalized Travelling wave
solutions of this equation are obtained with the help of symbolic
computation. From these results, we can see that the Exp-function
method is one of the most effective methods to obtain exact
solutions.\par Finally, it is worthwhile to mention that the
Exp-function method can be also extended to other nonlinear
evolution equations with variable coefficients, such as the mKdV
equation, the (3+1)-dimensional Burgers equation, the generalized
Zakharov-Kuznetsov equation and so on. The Exp-function method is a
promising and powerful new method for nonlinear evolution equations.
This is our task in future work.

\section*{ACKNOWLEDGMENTS}
In this work, the Maple library - Epsilon by Dr. Dongming Wang
is used to perform some of the computational tasks.
This work is funded by the National Basic Research Program of China
(973 Program No.2006CB705500), the National Natural Science
Foundation of China (Grant Nos. 60744003, 10635040, 10532060,
10472116), by the Special Research Funds for Theoretical Physics
Frontier Problems (NSFC No.10547004 and A0524701), by the President
Funding of Chinese Academy of Science, and by the Specialized
Research Fund for the Doctoral Program of Higher Education of China.


\begin{thebibliography}{99}
\bibitem{1} Abdusalam HA.On an improved complex tanh-function
method. Int J Nonlinear Sci Numer Simul 2005; 6(2):99-106.
\bibitem{2} Zayed EME, Zedan HA, Gepreel KA. Group analysis and
Modified extended tanh-function to fund the invariant solutions and
soliton solutions for nonlinear Euler equations. Int J Nonlinear Sci
Numer Simul 2004;5(3):221-34.
\bibitem{3}Bai CL, Zhao H. Generalized extended tanh-function method
and its application. Chaos, Solutions \& Fractals
2006;27(4):1026-35.
\bibitem{4}S.A.El-Wakil, M.A. Abdou. New exact travelling wave
solutions using modified extended tanh-function method. Chaos,
Solution \& Fract.2007;31(4):840-852.
\bibitem{5} E.Fan,.Extended tanh-function method and its
applications to nonlinear equations. Phys. Lett.A 2000;277:212-218.
\bibitem{6} Yomba E. The modified extended Fan sub-equation method
and its application to the (2+1)-dimensional Broer-Kaup-Kupershmidt
equation. Chaos ,Solitons \& Fractals 2006;27(1):187-96.
\bibitem{7} Ren YJ, Zhang HQ. A generalized F-expansion method to
find abundant families of Jacobi elliptic function solutions of the
(2+1)-dimensional Nizhnik-Novikov-Veselov equation. Chaos ,Solitons
\& Fractals 2006;27(4):959-79.
\bibitem{8} A.M.Wazwaz. Exact solutions to the double sinh-Gordon
equation by the separated ODE method. Comput. Math. Appl.
50(10-11)(2005)1685-1696.
\bibitem{9} C.Yan. A simple transformation for nonlinear waves,
Phys. Lett. A 1996;246:77-84.
\bibitem{10} A.M. Wazwaz . A sine-cosine method for handing nonlinear
wave equations. Math. Comput. Model.2004; 40:499-508.
\bibitem{11} Dai CQ, Zhang JF. Jacobian elliptic function method for
nonlinear differential-difference equations. Chaos, Solitions \&
Fractals 2006;27(4):1042-7.
\bibitem{12} Zhao XQ,Zhi HY,Zhang HQ. Improved Jacobi-function method
with symbolic computation to construct new double-periodic solutions
for the generalized Ito system. Chaos, Solitions \& Fractals
2006;28(1):112-26.
\bibitem{13} Yu Yx,Wang Q, Zhang HQ .The extended Jacobi ellipitic
function method to solve a generalized Hirota-Satsuma coupled KdV
equations . Chaos, Solitions \& Fractals 2005;26(5):1415-21.
\bibitem{14} E.Fan, J Zhang. Applications of the Jacobi elliptic
function method to special-type nonlinear equations. Phys. Lett. A
305(2002)383-392.
\bibitem{15} He JH, Application of homotopy perturbation method to
nonlinear wave equations. Chaos, Solitons \& Fractals
2005;26(3):695-700.
\bibitem{16} He JH. Homotopy perturbation method for bifurcation of
nonlinear problems. Int J Nonlinear Sci Numer Simul 2005;6(2):207-8.
\bibitem{17} El-Shahed M. Application of He's Homotopy perturbation
method to Volterra's integro-differential equation. Int J Nonlinear
Sci Numer Simul 2005;6(2):163-8.
\bibitem{18} J.H.He. New interpretation of  homotopy-perturbation
method . Int.J.Mod.Phys.B 2006;20(18):2561-2568.
\bibitem{19} J.H.He. Some asymptotic methods for strongly nonlinear
equations. Int.J.Mod.Phys.B 2006;20(10):1141-1199
\bibitem{20} J.H.He, X.H.Wu. Construction of solitary solution and
compacton-like solution by variational iteration method. Chaos,
Soliton \& Fractals.2006; 29(1):108-113
\bibitem{21} Abassy TA, El-Tawil MA, Saleh HK, The solution of KdV
and mKdV equations using Adomian Pade approximation. Int J Nonlinear
Sci Numer Simul 2004;5(4):327-39.
\bibitem{22} El-Sayed SM, Kaya D, Zarea S. The decomposition method
applied to solve high-order linear Volterra-Fredholm
integro-differential equations. Int J Nonlinear Sic Numer Simul
2004;5(2):105-12.
\bibitem{23} El-Danaf TS,Ramadan MA, Alaal FEIA. The use of adomian
decomposition method for solving the regularized long-wave equation
.Ghaos, Solitons \& Fractals 2005;26(3):747-57.
\bibitem{24}J.H.He,X.H.Wu. Exp-funtion method for nonlinear wave equations. Phys. Lett.
A 2006; 30:700-708.
\bibitem{25} J.H.He, M.A.Abdou. New periodic solutions for nonlinear
evolution equations using Exp-function method. Chaos Solitons \&
Fractals 2007;34(5):1421-1429.
\bibitem{26} Changbum Chun. Solitions and periodic solutions for
fifht-order KdV eqaution with the Exp-function method. Phys.Lett.A
2008; 372:2760-2766.
\bibitem{27} Alvaro H.Salas. Exact solutions for the general fifth
KdV equation by the exp function method. Appl.Math.Comput 2008;205:
291-297.
\bibitem{28} A. N. Kolmogorov, I. G. Petrovskii, and N. S. Piskunov. ¡°Investigation of the diffusion
equation with growth of the quantity of matter and its application
to a biology problem¡±, Byull. Moskov. Gos. Univ. Mat. Mekh. 1
(1937), no. 6, 1¨C26; English transl., Dynamics of curved fronts (P.
Pelc¡äe, ed.), Acad. Press, Boston 1988,pp 105¨C130.
\bibitem{29}M. J. Ablowitz and A. Zeppetella. Explicit solutions of Fisher's equation for a special wave speed. Bull. Math. Biol. 1979;41:835-840.
\bibitem{30}Jiang Lu, Guo Yu-Cui,Xu Shu-Jiang.Some new exact solutions to the Burgers¨CFisher
equation and generalized Burgers¨CFisher equation. Chin. Phys.
Soc.2007;16(09):2514-09.
\end{thebibliography}
\end{document}